\documentclass[11pt,epsf,letterpaper]{article}%
\usepackage{color}
\usepackage{amsmath}
\usepackage{amsfonts}
\usepackage{verbatim}
\usepackage{amssymb}
\usepackage{graphicx}
\usepackage{epstopdf}
\usepackage{mathrsfs}%
\providecommand{\U}[1]{\protect\rule{.1in}{.1in}}
\begin{document}

\title{Black holes in $\omega$-defomed gauged $N=8$ supergravity}
\author{$^{(1,2)}$Andr\'{e}s Anabal\'{o}n and $^{(3)}$Dumitru Astefanesei.\\\textit{$^{(1)}$Departamento de Ciencias, Facultad de Artes Liberales y}\\\textit{Facultad de Ingenier\'{\i}a y Ciencias, Universidad Adolfo
Ib\'{a}\~{n}ez,}\\\textit{Av. Padre Hurtado 750, Vi\~{n}a del Mar, Chile.}\\\textit{$^{(2)}$ Universit\'{e} de Lyon, Laboratoire de Physique, UMR 5672,
CNRS,}\\\textit{\'{E}cole Normale Sup\'{e}rieure de Lyon}\\\textit{46 all\'{e} d'Italie, F-69364 Lyon Cedex 07, France.}\\\textit{$^{(3)}$Instituto de F\'\i sica, Pontificia Universidad Cat\'olica de
Valpara\'\i so,} \\\textit{ Casilla 4059, Valpara\'{\i}so, Chile.} \\andres.anabalon@uai.cl, dumitru.astefanesei@ucv.cl}
\maketitle

\begin{abstract}
Motivated by the recently found $4$-dimensional $\omega$-deformed gauged
supergravity, we investigate the black hole solutions within the single
scalar field consistent truncations of this theory. We construct black hole
solutions that have spherical, toroidal, and hyperbolic horizon topology. The
scalar field is regular everywhere outside the curvature singularity and the
stress-energy tensor satisfies the null energy condition. When the parameter
$\omega$ does not vanish, there is a degeneracy in the spectrum of black hole
solutions for boundary conditions that preserve the asymptotic Anti-de Sitter
symmetries. These boundary conditions correspond to multi-trace deformations
in the dual field theory.

\end{abstract}

\section{Introduction}

Supergravity is an effective theory for string theory at low energies compared
to the string energy scale. The Anti-de Sitter/Conformal Field Theory
(AdS/CFT) duality \cite{Maldacena:1997re} provides a remarkable tool for
extracting information about strongly coupled gauge theories (in $d$
dimensions) from a dual supergravity description (in $d+1$ dimensions). Within
the AdS/CFT duality, the radial coordinate plays the role of the energy scale
and so the bulk geometry has a nice interpretation as a renormalization group
(RG) flow of the dual field theory \cite{Akhmedov:1998vf} (see, also, the
review \cite{Skenderis:2002wp}).

The gauged $N=8$ supergravity in four dimensions is the maximal gauged
supergravity with spins lower or equal than two and it can be obtained by
Kaluza-Klein (KK) reduction of $11$-dimensional supergravity
\cite{Cremmer:1978km} on a $7$-dimensional sphere \cite{de Wit:1986iy}. In the
holographic context, much of the interest on the $4$-dimensional gauged
supergravities comes from the utility of the ABJM model \cite{Aharony:2008ug}
in testing various strongly coupled phenomena in condensed matter physics.

What came as a surprise, recently, is the existence of a continuous
one-parameter family of inequivalent maximally supersymmetric gauged
supergravities \cite{Dall'Agata:2012bb} (the critical points were 
extensively studied in \cite{Borghese:2012qm}). These theories are characterized by
one `angular parameter', $\omega$, that generates an electric-magnetic duality
transformation prior to selecting the $SO(8)$ gauging, and they are referred
to as $\omega$-deformed gauged $N=8$ supergravity. At the practical level, the
$\omega$ parameter can be introduced in the maximal gauged supergravity by
multiplying the $56$-bein, $\mathcal{V}$, by a diagonal matrix containing
either $e^{i\omega}$ or $e^{-i\omega}$ \cite{deWit:2013ija}. As the moduli
potential is a non-linear function in terms of $\mathcal{V},$ the parameter
$\omega$ appears explicitly in the potential. It is worth noting that, long
time ago, this kind of `angles' were introduced in $N=4$ gauged supergravities
in \cite{deRoo:1985jh}.

In this Letter, we construct \textit{exact} domain wall (see, also, \cite{Guarino:2013gsa}) 
and black hole solutions in one scalar field consistent truncations of $\omega$-defomed
theories \cite{Tarrio:2013qga}, which correspond to RG flows of 3-dimensional
dual quantum field theories at zero and finite temperatures, respectively. We
show that the null energy condition (which is the relevant energy condition in
AdS) is satisfied and construct the c-function using the gravity side of the duality.

For $\omega=0$, we obtain domain wall solutions and explicitly write down the
corresponding superpotential. When $\omega$ does not vanish, there is a
degeneracy in the spectrum of black hole solutions. Since the black holes are
interpreted as thermal states in the dual theory, this degeneracy may seem
puzzling at first sight. However, this can be understood because the solutions
correspond to different boundary conditions which in turn correspond to
different deformations of the dual field theory. The degeneracy in the
spectrum of solutions can then be understood as a sign degeneracy in the AdS
invariant boundary conditions due to a change in the sign of the scalar field.

Since the coordinates in which the black hole solutions are constructed are
not very intuitive, we will explicitly provide the change of coordinates that
asymptotically puts the metric in the canonical form \cite{Hertog:2004dr}. In
this system of coordinates, it is easy to observe that the scalar field
satisfies mixed boundary conditions that correspond to multi-trace
deformations in the dual theory \cite{Witten:2001ua}.

\section{Moduli potential}

The single scalar field truncations of the $\omega-$deformed theory were found
in \cite{Tarrio:2013qga}. The action is the one of Einstein gravity minimally
coupled to a scalar field
\begin{equation}
I[g_{\mu\nu},\varphi]=\frac{1}{16 \pi G}\int d^{4}x\sqrt{-g}\left[  R-\frac
{1}{2}\partial_{\mu}\varphi\partial^{\mu}\varphi-V_{n}(\varphi)\right]
\label{I}%
\end{equation}
where $G$ is the Newton constant. The moduli potentials are parameterized by
an integer, $n=\{1..7\}$, and they contain a non-trivial deformation
parameter, $\omega$, when $n$ is odd \cite{Tarrio:2013qga}.\footnote{$V_{1}%
(\varphi)$ was also discussed and independently derived in
\cite{deWit:2013ija}.}

We found that all the truncations are contained in a more general moduli
potential%
\begin{align}
W_{\nu}(\varphi)  &  =\frac{2\alpha}{\nu^{2}}\left[  \frac{\nu-1}{\nu+2}%
\sinh(\varphi l_{\nu}\left(  \nu+1\right)  )-\frac{\nu+1}{\nu-2}\sinh(\varphi
l_{\nu}\left(  \nu-1\right)  )+4\frac{\nu^{2}-1}{\nu^{2}-4}\sinh(\varphi
l_{\nu})\right] \nonumber\\
&  -\frac{\left(  \nu^{2}-4\right)  }{l^{2}\nu^{2}}\left[  \frac{\nu-1}{\nu
+2}\exp(-\varphi l_{\nu}\left(  \nu+1\right)  )+\frac{\nu+1}{\nu-2}%
\exp(\varphi l_{\nu}\left(  \nu-1\right)  )+4\frac{\nu^{2}-1}{\nu^{2}-4}%
\exp(-\varphi l_{\nu})\right]  \label{W}%
\end{align}
where $l$ is the AdS radius and $l_{\nu}^{-2}=(\nu^{2}-1)$. To obtain the
potentials $V_{n}(\varphi)$ from the general potential $W_{\nu}(\varphi)$, we
have to relate the parameters $\omega$ and $g$ to $\alpha$ and the AdS radius,
$l$, in the following way: $\sin(\omega)^{2}=\frac{\alpha S_{n}}{g^{2}}$,
$g=\frac{1}{\sqrt{2}l}$, where for odd $n$ we define $S_{1}=\frac{9}{40}$,
$S_{3}=-\frac{1}{24}$, $S_{5}=\frac{1}{24}+\frac{g^{2}}{\alpha}$, and
$S_{7}=-\frac{9}{40}+\frac{g^{2}}{\alpha}$. Then, by using the relations
between parameters, it is straightforward to show that
\begin{align}
V_{1}(\varphi)  &  =W_{\frac{4}{3}}(\varphi)=V_{7}(-\varphi)\qquad,\qquad
V_{2}(\varphi)=\left.  W_{2}(\varphi)\right\vert _{\alpha=0}=V_{6}%
(\varphi)\label{C1}\\
V_{3}(\varphi)  &  =W_{4}(\varphi)=V_{5}(-\varphi)\qquad,\qquad V_{4}%
(\varphi)=\left.  W_{\infty}(\varphi)\right\vert _{\alpha=0} \label{C2}%
\end{align}

The scalar field potential (\ref{W}) has a negative maximum and so AdS is a
solution.\footnote{Obviously, this is the critical point that preserves $N=8$ 
supersymmetry and the group $SO(8)$, however there exist also additional 
non-supersymmetric critical points \cite{Borghese:2012qm}.} More precisely, 
the AdS vacuum can be obtained for $\varphi=0$
\begin{equation}
W_{\nu}(\varphi)=-\frac{3}{l^{2}}\qquad,\qquad\frac{dW_{\nu}(\varphi
)}{d\varphi}=0\qquad,\qquad\frac{d^{2}W_{\nu}(\varphi)}{d\varphi^{2}}%
=-\frac{2}{l^{2}}%
\end{equation}
Indeed, the small scalar fluctuations are tachyonic, $m^{2}<0$, but the mass
is still above the Breitenlohner-Freedman (BF) bound, $m_{BF}^{2}=-\frac
{9}{4l^{2}}$ and so this vacuum is perturbatively stable \cite{BF}. Moreover,
since the mass is in the range $m_{BF}^{2}<m^{2}<m_{BF}^{2}+l^{-2}$, both
modes of the scalar field are normalizable \cite{Ishibashi:2004wx}. Therefore,
these theories are suitable in the context of AdS/CFT duality with boundary
conditions corresponding to the presence of multi-trace operators in the dual
field theory \cite{Witten:2001ua}.

We would like to point out that it can be shown that for these one scalar
field consistent truncations, the range of the parameter is $\omega\in
\lbrack0,\frac{\pi}{4}]$ \cite{Tarrio:2013qga} rather than $\omega\in
\lbrack0,\frac{\pi}{8}]$ like in the case of the full theory
\cite{Dall'Agata:2012bb}. The reason behind this fact is that the fields for
which the identifications should be done in order to reduce the range of
$\omega$ have been truncated out.

\section{Exact black hole solutions}

In this section, we investigate the existence of event horizons in terms of
the parameters of the potential. Similar techniques to find hairy black holes
were used in \cite{Anabalon:2012ta, Acena:2012mr, Anabalon:2013qua} (see,
also, \cite{Anabalon:2013sra, Acena:2013jya}) where the details of the
construction can be found. In what follows, we present the solutions, discuss
some of their properties, and explicitly check the existence of the horizon.

The black holes in AdS can have non-spherical horizon topology and the
solutions relevant to $\omega$-deformed gauge supergravity can be written in a
compact form as
\begin{align}
ds^{2}  &  =\Omega(x)\left[  -f(x)dt^{2}+\frac{\eta^{2}dx^{2}}{f(x)}%
+d\Sigma_{k}\right] \label{S1}\\
\varphi &  =l_{\nu}^{-1}\ln(x)\qquad,\qquad\Omega(x)=\frac{\nu^{2}x^{\nu-1}%
}{\eta^{2}\left(  x^{\nu}-1\right)  ^{2}} \label{O}%
\end{align}%
\begin{equation}
f(x)=\frac{x^{2-\nu}\left(  x^{\nu}-1\right)  ^{2}\eta^{2}k}{\nu^{2}}+\left[
\frac{1}{\nu^{2}-4}-\frac{x^{2}}{\nu^{2}}\left(  1+\frac{x^{-\nu}}{\nu
-2}-\frac{x^{\nu}}{\nu+2}\right)  \right]  \alpha+\frac{1}{l^{2}} \label{S2}%
\end{equation}
Here, $d\Sigma_{k}$ is a two-manifold of constant curvature $k=\pm1$ or $0$
and the corresponding black holes can have spherical, toroidal or hyperbolic
horizon topologies. The parameter $\eta$ is in fact the integration constant
and the reason it appears in a non-standard form in the metric is because the
computations simplify when a dimensionless radial coordinate, $x$, is used.

The configuration and scalar potential are invariant under the change
$\nu\rightarrow-\nu$, therefore from now on we consider $\nu\geq1$. We also
note that conformal infinity is at $x=1$. There are two branches of solutions
depending whether $x\in(0,1)$ or $x\in(1,\infty)$. The metric is regular for
any value of $x\neq0$ and $x\neq\infty$ as can be seen from the introduction
of advanced and retarded coordinates $u_{\pm}=t\mp\int\frac{\eta}{f(x)}dx$.
The scalar field and metric are singular at $x=0$ and $x=\infty$ but, as we
will see, these singularities can be covered by event horizons.

Interestingly, it is straightforward to observe in our parameterization that,
with this scalar field potential, the theory also admits asymptotically flat
hairy black holes \cite{Anabalon:2012ih, Anabalon:2013qua, Anabalon:2013baa}
(when $l^{-2}=0$) that could, in principle, be related to asymptotically flat
gauged supergravity \cite{Sezgin}. Solutions of the form (\ref{S1})-(\ref{S2})
in four dimensions were originally found in \cite{Anabalon:2012ta} and
discussed in different contexts in the literature, \cite{Salvio:2012at,
Salvio:2013ja, Salvio:2013jia, Anabalon:2013sra}. Indeed, these solutions
exist for any value of $\nu\,$and $\alpha$ although we shall focus below only
on the values related to gauged supergravity.

For even $n$ and vanishing $\alpha$ the situation is well known and has been
thoroughly studied. When $n$ is even, there exist exact neutral asymptotically
locally AdS black holes only for $k=-1$ \cite{Anabalon:2012sn}. In what
follows, we are going to carefully analyze the existence of event horizons for
$k=1$ and comment also on the case $k=0$. Since the solutions are static, the
horizon is localized at the place where the $g_{tt}$ component of the metric
vanishes. That is, there is an $x_{+}$ such that $f(x_{+})=0$. As expected,
since the solution is asymptotically AdS, we obtain at the boundary, where the
scalar field vanishes, $f(x=1)=l^{-2}$. Furthermore, it is straightforward to
obtain the (non-trivial) solution of the equation $\frac{df}{dx}=0$, and by
analyzing the second derivative of $f$ we obtain that this point is a maximum.
Therefore, the function $f$ has at most one zero in any physically relevant
interval. Since $f$ is positive at the boundary, the existence of a black hole
horizon is ensured if $f$ is negative near the singularities. Therefore, the
necessary and sufficient condition for the existence of a horizon are
\begin{align}
\nu &  <2\text{, }x<1\Longrightarrow\frac{\alpha}{\nu^{2}-4}+\frac{1}{l^{2}%
}<0\qquad\nu<2\text{, }x>1\Longrightarrow\left(  \eta^{2}+\frac{\alpha}{\nu
+2}\right)  <0\label{A}\\
\nu &  >2\text{, }x<1\Longrightarrow\left(  \eta^{2}-\frac{\alpha}{\nu
-2}\right)  <0\qquad\nu>2\text{, }x>1\Longrightarrow\left(  \eta^{2}%
+\frac{\alpha}{\nu+2}\right)  <0 \label{B}%
\end{align}
Therefore, there should exist hairy black holes in an open set of the
parameter space when at least one of the relations above are satisfied.

Finally, we want to point out that the symmetries of these theories (\ref{C1})
and (\ref{C2}) allows to map the $n=3$ into the $n=5$ theory by making the
field redefinition $\varphi\rightarrow-\varphi$ and letting the geometry invariant.

\subsection{Spherically symmetric black holes}

We are going to consider the cases of non-trivial $\alpha$, namely when $n$ is odd.

\begin{itemize}
\item When $n=1$ then $\alpha=\frac{\sin(\omega)^{2}}{l^{2}}\frac{20}{9}$ and
$\nu=\frac{4}{3}$. Then, the condition (\ref{A}) shows that there are no
spherically symmetric black holes.

\item When $n=3$ then $\alpha=-\frac{12\sin(\omega)^{2}}{l^{2}}$ and $\nu=4$.
Then, the condition (\ref{B}) shows that there are black holes when $x>1$ and
$\sin(\omega)^{2}>\frac{l^{2}\eta^{2}}{2}$.

\item When $n=5$ then $\sin(\omega)^{2}=\frac{\alpha l^{2}}{12}%
+1\Longrightarrow\cos(\omega)^{2}=-\frac{\alpha l^{2}}{12}$ and $\nu=4$. Then,
the condition (\ref{B}) shows that there are black holes when $x>1$ and
$\cos(\omega)^{2}>\frac{l^{2}\eta^{2}}{2}$.

\item When $n=7$ then $\sin(\omega)^{2}=-\frac{9\alpha l^{2}}{20}%
+1\Longrightarrow\cos(\omega)^{2}=\frac{9\alpha l^{2}}{20}$ and $\nu=\frac
{4}{3}$. Then, the condition (\ref{A}) shows that there are no spherically
symmetric black holes.
\end{itemize}

\subsection{Planar black holes}

For $k=0$, the condition that $f(x)$ becomes negative near the singularity is
\begin{align}
\nu &  <2\text{, }x<1\Longrightarrow\frac{\alpha}{\nu^{2}-4}+\frac{1}{l^{2}%
}<0\qquad\nu<2\text{, }x>1\Longrightarrow\frac{\alpha}{\nu+2}<0\\
\nu &  >2\text{, }x<1\Longrightarrow-\frac{\alpha}{\nu-2}<0\qquad\nu>2\text{,
}x>1\Longrightarrow\frac{\alpha}{\nu+2}<0
\end{align}
As in the previous case, there are no black holes when $n=1$ or $7$.

\begin{itemize}
\item When $n=3$ then condition (\ref{B}) shows that there are black holes
when $x>1$ and for $\forall\omega\neq0$.

\item When $n=5$ then condition (\ref{B}) shows that there are black holes
when $x>1$ and for $\forall\omega$.
\end{itemize}

When $k=0$ and $\alpha=0$, we also obtain domain wall solutions:
\begin{align}
ds^{2}  &  =l^{2}\Omega(x)\left[  -dt^{2}+\eta^{2}dx^{2}+dy^{2}+dz^{2}\right]
\\
\varphi &  =l_{\nu}^{-1}\ln(x)\qquad,\qquad\Omega(x)=\frac{\nu^{2}x^{\nu-1}%
}{\eta^{2}\left(  x^{\nu}-1\right)  ^{2}}%
\end{align}
In this case the potential can be explicitly written in terms of a
superpotential
\begin{equation}
W_{\nu}(\varphi)=2\left(  \frac{dP(\varphi)}{d\varphi}\right)  ^{2}-\frac
{3}{2}P(\varphi)^{2}%
\end{equation}
where
\begin{equation}
P(\varphi)=l^{-1}\left[  \frac{\left(  \nu+1\right)  }{\nu}e^{\left(
\frac{\nu-1}{2}\right)  \varphi l_{\nu}}+\frac{\left(  \nu-1\right)  }{\nu
}e^{-\left(  \frac{\nu+1}{2}\right)  \varphi l_{\nu}}\right]
\end{equation}

\section{Boundary conditions}

One important question we would like to address in this section is if the
boundary conditions for our solutions are compatible with the AdS/CFT duality.
The mixed boundary conditions compatible with AdS/CFT duality were interpreted
as a multi-trace deformation of the boundary CFT \cite{Witten:2001ua}.

Let us discuss the branch $x\in(1,\infty)$ for which the scalar field is
positively defined. We change the coordinates so that the function in front of
the transversal section, $d\Sigma_{k},$ has the following fall-off:
\begin{equation}
\Omega(x)=r^{2}+O(r^{-3}) \label{CC}%
\end{equation}
By using the change of coordinates\footnote{We assume that $\eta>0$ so that
$r\rightarrow\infty$ implies $x\rightarrow1$ from the right.}
\begin{equation}
x=1+\frac{1}{\eta r}-\frac{(\nu^{2}-1)}{24\eta^{3}r^{3}}\left[  1-\frac
{1}{\eta r}-\frac{9\left(  \nu^{2}-9\right)  }{80\eta^{2}r^{2}}\right]
\end{equation}
we obtain the following asymptotic expansions for the metric functions:
\begin{equation}
-g_{tt}=f(x)\Omega(x)=\frac{r^{2}}{l^{2}}+k+\frac{\alpha+3\eta^{2}k}{3\eta
^{3}r}+O(r^{-3}) \label{Lapse}%
\end{equation}%
\begin{equation}
g_{rr}=\frac{\Omega(x)\eta^{2}}{f(x)}\left(  \frac{dx}{dr}\right)  ^{2}%
=\frac{l^{2}}{r^{2}}-\frac{l^{4}}{r^{4}}-\frac{l^{2}(\nu^{2}-1)}{4\eta
^{2}r^{4}}+\frac{l^{2}\left(  3kl^{2}\eta^{2}+l^{2}\alpha-\nu^{2}+1\right)
}{3\eta^{3}r^{5}}+O(r^{-6}) \label{grr}%
\end{equation}
Therefore, the asymptotic behavior of our exact black hole solutions fits very
well in the general analysis of \cite{Hertog:2004dr} and, consequently, we
expect that the boundary conditions for the scalar correspond to a multi-trace
deformation in the dual field theory \cite{Witten:2001ua}. The asymptotic
expansion of the scalar field is
\begin{equation}
l_{\nu}\varphi=\frac{1}{\eta r}-\frac{1}{2\eta^{2}r^{2}}-\frac{\nu^{2}%
-9}{24\eta^{3}r^{3}}+O(r^{-4}) \label{expan}%
\end{equation}
and by comparing with the scalar field in the standard notation of
\cite{Hertog:2004dr}
\begin{equation}
\varphi=\frac{a}{r^{\lambda_{-}}}+\frac{b}{r^{\lambda+}}%
+...\,\,\,\,\,\,\,\,\,\,\lambda_{-}<\lambda_{+} \label{con}%
\end{equation}
we obtain $\lambda_{+}=2\lambda_{-}=2$, $a=\frac{1}{\eta l_{\nu}}$, and
$b=-\frac{1}{2\eta^{2}l_{\nu}}$. The scalar field introduces an asymptotic
deviation at the order $O(r^{-4})$ in (\ref{grr}) that is not present when the
scalar field vanishes ($\nu^{2}=1$).

Since $b=-\frac{l_{\nu}}{2}a^{2}$, these boundary conditions are, indeed,
compatible with the canonical realization of the conformal symmetry at the
boundary. However, when the scalar field is negative, the boundary conditions
change and are of the form $b=\frac{l_{\nu}}{2}a^{2}$ and there is a
degeneracy in the spectrum of black hole solutions. These boundary conditions
are invariant under the asymptotic AdS symmetries. Black holes are interpreted
as thermal states in the dual field theory, and from this point of view, $b$
is the expectation value of an operator of dimension two. The degeneracy is
absent when $\omega=0$ because the black holes with $n=3$ and $\varphi>0$ do
not exist in this case. It is also worth to remark that AdS invariant
multi-trace deformation are not incompatible with supersymmetry, as was
checked in some concrete examples in \cite{Duff:1999gh}.

\section{Discussion}

In this Letter, we have investigated the existence of \textit{exact} domain
wall and black hole solutions in one scalar field consistent truncations of
$\omega$-deformed $N=8$ sugra theories. The black holes only exist in the
$n=3$ and $n=5$ theory, which are actually identical up to the field
redefinition $\varphi\longrightarrow-\varphi$. Let us therefore focus on the
theory with $n=5$. This theory has a hairy black hole with AdS invariant
boundary conditions even when $\omega=0$ (because $\alpha$ does not vanish in
this case). The value of the scalar field is everywhere positive as it has no
nodes. By turning on the deformation parameter, the theory has a new black
hole solution that corresponds to a negative scalar field, $\varphi<0$. From
our discussion in Section $4$, it is clear that this is equivalent to the fact
that there are black holes for boundary conditions of the form $b=\pm
\frac{l_{\nu}}{2}a^{2}$. Therefore, when $\omega\neq0$ there are \textit{two}
hairy black holes for each value of the parameters. Each hairy black hole is
associated to a sign flip in the expectation value of a dimension $2$ operator
in the CFT. We would also like to remark that, to the best of our knowledge,
the exact form of the black holes in undeformed gauged supergravity was unkown
until now.

The existence of these black holes also implies that there is a richer phase
diagram in the dual theory. Hairy black holes in $AdS_{4}$ are physically
relevant objects as they are an important ingredient of the AdS/Condensed
Matter correspondence (see, e.g, \cite{Hartnoll:2009sz} and the reference
therein). As was originally pointed out in \cite{Martinez:2004nb}, there could
also exist second order phase transitions between the hairy black holes and
the hairless ones.

For completeness of the analysis, let us now show that the null energy
condition is satisfied. Indeed, whenever there is a single minimally coupled
scalar field to a diagonal metric it is possible to show that the null energy
condition is satisfied. In an orthonormal frame $e_{\mu}^{a}$, the energy
momentum tensor has the form $T_{\mu\nu}e_{a}^{\mu}e_{b}^{\nu}=diag(\rho
,p_{1},p_{2},p_{2})$ where
\begin{equation}
\rho+p_{1}=g_{xx}^{-1}\left(  \frac{d\varphi}{dx}\right)  ^{2}\geq0,\qquad
\rho+p_{2}=0.
\end{equation}
Since the null energy condition is valid for our solutions, there are no
violations that lead to superluminal propagation and instabilities in the bulk
\cite{Gao:2000ga} and so we expect that the boundary theory is well defined
(see, e.g., \cite{Kleban:2001nh}). Also, the c-function
\cite{Freedman:1999gp} can be also constructed:\footnote{Similar examples are
presented in \cite{Astefanesei:2007vh, Anabalon:2013sra} and we do not present the details
here.}
\begin{equation}
C(x)=C_{0}\left(  \frac{2\eta\Omega^{3/2}}{\Omega^{\prime}}\right)
^{2}=\mathcal{C}_{0}\left[  \frac{x^{\frac{\nu+1}{2}}}{x^{\nu}(\nu+1)+(\nu
-1)}\right]  ^{2} \label{C11}%
\end{equation}
As expected, in the hairless limit when the neutral spherical black hole is
obtained ($\nu=1$), the flow is trivial. We also notice that the c-function is
completely determined by the conformal factor that is in agreement with the
interpretation of the black hole as a thermal state in the boundary theory.
The c-function should remain unchanged when a finite temperature vacuum is
excited in the same theory and so there should not be any dependence of the
metric function $f(x)$.

In conclusion, the black hole solutions we have found correspond to mixed boundary
conditions in AdS and can be interpreted within the AdS/CFT duality as RG
flows of the dual field theory at finite temperature.

\section{Acknowledgments}

We would like to thank Laura Andrianopoli and Henning Samtleben for
interesting discussions. Research of A.A. is supported in part by the Fondecyt
Grant 11121187 and by the CNRS project \textquotedblleft Solutions exactes en
pr\'{e}sence de champ scalaire\textquotedblright. The work of D.A. is
supported in part by the Fondecyt Grant 1120446.

\end{document}